# A comparative study of the thermoelectric performance of graphene-like BX (X= P, As, Sb) monolayers

Z. Z. Zhou, H. J. Liu,[*] D. D. Fan, G. H. Cao

*Key Laboratory of Artificial Micro- and Nano-Structures of Ministry of Education and School of Physics and Technology, Wuhan University, Wuhan 430072, China*

The electronic and phonon transport properties of graphene-like boron phosphide (BP), boron arsenide (BAs), and boron antimonide (BSb) monolayers are investigated using first-principles calculations and Boltzmann theory. By considering both the phonon-phonon and electron-phonon scatterings, we demonstrate that the strong bond anharmonicity in the BAs and BSb monolayers can dramatically suppress the phonon relaxation time but hardly affects that of electrons. As a consequence, both systems exhibit comparable power factors with that of the BP monolayer but much lower lattice thermal conductivities. Accordingly, a maximum *ZT* values above 3.0 can be achieved in both BAs and BSb monolayers at optimized carrier concentrations. Interestingly, very similar *p*- and *n*-type thermoelectric performance is observed in the BSb monolayer along the armchair direction, which is of vital importance in the fabrication of thermoelectric modules with comparable efficiencies.

Thermoelectric materials, which can directly convert waste heat into electricity or vise versa, have attracted much attention since most of the energy produced from fossil fuels is lost as waste heat.[1–3] Materials designed for thermoelectric applications should have suitable band gap, high carrier mobility, and low thermal conductivity. The pioneering work of Hicks and Dresselhaus[4,5] suggested that low-dimensional structures may achieve high thermoelectric performance compared with the bulk counterparts. As the best-known low-dimensional materials, the graphene[6–10] is however not suitable for thermoelectric applications due to the gapless energy band. To open a gap in the graphene, several methods such as hydrogenation[11,12] and fluorination[13,14] have been suggested. On the other hand, Sahin *et al.*[15] theoretically predicted that the graphene-like III-V binary compounds exhibit intrinsic band gaps. In particular, the boron phosphide (BP), boron arsenide (BAs), and boron antimonide (BSb) monolayers exhibit direct band gaps around 1.0 eV, which are very desirable as thermoelectric materials.

---

[*] Author to whom correspondence should be addressed. Electronic mail: phlhj@whu.edu.cn

In this study, we give a comparative study of the thermoelectric transport properties of the graphene-like BX (X= P, As, Sb) monolayers. The lattice thermal conductivity ($\kappa_l$) is computed by solving the phonon Boltzmann transport equation, as implemented in the so-called ShengBTE package.[16] The interatomic force constants are obtained by density functional theory (DFT) and the finite displacement method, as performed in the Vienna *ab-initio* simulation package (VASP)[17] and the PHONOPY program,[18] respectively. The band structures of the BX monolayers are calculated within the framework of DFT, which is implemented in the QUANTUM ESPRESSO package[19] where the core-valence interaction is described by the norm-conserving scalar-relativistic pseudopotentials.[20] The hybrid density functional in the form of Heyd-Scuseria-Ernzerhof (HSE)[21] is adopted to obtain accurate results, which has been successfully used to predict the band gap of the graphene-like BN monolayer.[22] The electronic transport coefficients, including the Seebeck coefficient ($S$), the electrical conductivity ($\sigma$), and the electronic thermal conductivity ($\kappa_e$), can be derived from the Boltzmann transport theory.[23] By fully considering the electron-phonon coupling (EPC), the ***k***-dependent carrier relaxation time ($\tau_c$) is calculated using the density functional perturbation theory (DFPT)[24] and the Wannier interpolation techniques,[25] which is coded in the electron-phonon Wannier (EPW) package.[26] The vacuum distance of 30 Å is adopted to simulate the monolayer structure, and all the calculated transport coefficients are renormalized with respect to the interlayer distance of graphite (3.35 Å).

Single-layer BX is *sp*$^2$ bonded with the B and X atoms arranged in a hexagonal honeycomb lattice. Similar to that of graphene, the primitive cell of BX monolayer contains two atoms, and each B atom is three-fold coordinated to the neighboring X atoms. The calculated B−X bond lengths of BP, BAs, and BSb monolayers are 1.86 Å, 1.96 Å, and 2.16 Å, respectively, which are in good agreement with previous theoretical results.[27] As there is no imaginary frequency in the phonon dispersion relations shown in Figure 1(a)-1(c), the dynamic stability of these BX monolayers is guaranteed. Besides, the in-plane Young's modulus (141.9 N/m, 121.5 N/m, and 90.8 N/m for the BP, BAs, and BSb, respectively) are comparable with that of the MoS$_2$ monolayer[28] and obviously larger than that of the phosphorene,[29] indicating the mechanical stability of these systems. Moreover, we see from the insets of Fig. 1(a)-1(c) that the B-X distance varies slightly around the equilibrium bond length during the *ab-initio* molecular dynamics (AIMD) simulation at 1400 K, suggesting the high temperature stability of the BX series. Similar to those found in graphene[30] and hexagonal BN,[31] we see that all the optical branches of BX monolayers exhibit very weak dispersions, which suggests that their lattice thermal conductivities should be mainly contributed by the acoustic phonons. With increasing atomic mass of the X elements, we find that the maximum phonon frequency becomes smaller, indicating the gradually lower lattice thermal conductivity.[32] Fig. 1(d)-1(f) plot the lattice thermal conductivities of BP, BAs, and BSb monolayers as a function of temperature, respectively. For each system, we see that the lattice thermal conductivity along the zigzag direction is slightly lower than that along the armchair direction. At room temperature, the $\kappa_l$ of BP, BAs, and BSb monolayers are 240 (312) W/mK, 27.6 (36.3) W/mK, and 18.7 (26.3) W/mK along the zigzag (armchair) direction, respectively. It is clear that $\kappa_l$ decreases by one order of magnitude from BP to BAs and BSb monolayers. To have a deep understanding, we compare the acoustic phonon group velocities ($v_{ph}$) of BX monolayers at 300 K. As displayed in Table I, the $v_{ph}$ of each acoustic phonon mode in the BAs and BSb monolayers are obviously smaller than that in the BP systems. Such an observation suggests much lower lattice thermal conductivities of BAs and BSb monolayers since $\kappa_l \propto v_{ph}^2$. The decreasing phonon group velocities from BP, BAs to BSb are consistent with their reducing highest acoustic phonon frequencies (see Fig. 1(a)-1(c)).[33] Note that several previous works suggest that the ZA modes contribute most to the lattice thermal conductivity of graphene-like structures owing to the selection rules.[34–36] Here we find that the group velocities of ZA modes for BAs and BSb monolayers are close to those of the good thermoelectric materials with intrinsic low $\kappa_l$, such as $Bi_2Te_3$[37] and PbTe.[38] Moreover, we see from the insets of Fig. 1(d)-1(f) that the phonon relaxation time ($\tau_{ph}$) of the acoustic modes in the BP monolayer is dramatically larger than those of the BAs and BSb systems, which can be attributed to the weaker anharmonicity of the B−P bond.[39] Indeed, compared with that of the BP monolayer (1.36), the absolute value of the Grüneisen parameters (γ) of the BAs (2.94) and BSb (2.75) systems are much higher due to the bigger mass difference of the B and X atoms.[40] Note that such higher γ values are comparable to those of bulk SnSe[41] and $Bi_2Te_3$.[40] It should be mentioned that although the γ of the BAs monolayer is relatively larger than that of the BSb, the larger $v_{ph}$ of the former compensates its stronger anharmonicity and consequently leads to a relatively higher lattice thermal conductivity of BAs system.

Figures 2(a)-2(c) plot the HSE band structures of BP, BAs, and BSb monolayers, respectively. Similar to those of the graphene[9] and hexagonal BN,[15] the conduction band minimum (CBM) and valance band maximum (VBM) of BX series are both located at the **K** point. The direct band gaps of BP, BAs, and BSb monolayers are 1.37 eV, 1.18 eV, and 0.61 eV, respectively, which decrease with increasing atomic mass of the X elements. As can be found in the inset of Fig. 2(c), the CBM and VBM of BSb monolayer are mainly contributed by the $p_z$ orbitals of the B and Sb atoms, respectively. Similar pictures can be found in the BP and BAs monolayers, indicating that the band gaps of the BX series are determined by the energy difference between the $p_z$ orbitals of the B and X atoms ($\Delta E_{p_z}$). From BP to BAs and BSb monolayers, the $\Delta E_{p_z}$ decreases and is responsible for the reduced band gaps. Our calculated results are consistent with previous study using a tight binding analysis.[27] On the other hand, it can be found that the energy dispersions around the VBM and CBM become stronger when going from the BP to BAs and BSb monolayers, which means

gradually lower density of state effective masses ($m^*_{dos}$) of the series. Indeed, the $m^*_{dos}$ of hole (electron) for the BP, BAs, and BSb monolayers are calculated to be 0.183 (0.195) $m_e$, 0.165 (0.173) $m_e$, and 0.105 (0.106) $m_e$, respectively. Such an observation suggests that higher carrier mobility and thus larger electrical conductivity could be found in the BSb monolayer. Besides, the almost identical $m^*_{dos}$ of hole and electron in the BSb monolayer indicates that the electronic transport properties of *p*- and *n*-type system may be comparable, which is very desirable for thermoelectric modules. Fig. 2(d) displays the energy-dependent carrier relaxation time of BP, BAs, and BSb monolayers at 300 K. For both *p*- and *n*-type systems, it is obvious that the relaxation time of BSb around the band edges is relatively higher than those of BP and BAs monolayers. This is reasonable since the BX series exhibit similar deformation potential constants (−2.3 eV and −3.8 eV for the *p*- and *n*-type systems, respectively) but distinct $m^*_{dos}$ mentioned above. Such finding also suggests that although the strong anharmonicity in the BAs and BSb monolayers can greatly suppress the $\tau_{ph}$ and $\kappa_l$, it hardly affects the $\tau_c$ and the electronic transport.

Using the Boltzmann transport theory and inserting the above-mentioned ***k***-dependent carrier relaxation time, we can evaluate the electronic transport properties of the BX monolayers. As an example, we list in Table II the room temperature $S$, $\sigma$, $\kappa_e$, power factor ($S^2\sigma$), and carrier mobility ($\mu$) of *p*-type BX series along the zigzag direction. It is obvious that $S^2\sigma$ of the BSb and BAs monolayers are much higher than that of the BP system, which is attributed to the simultaneously larger Seebeck coefficient and electrical conductivity originated from the lower carrier concentration and smaller $m^*_{dos}$, respectively. Compared with that of good thermoelectric material SnSe,[42] the BX monolayers show much larger power factors due to their ultrahigh carrier mobilities (even comparable with that of graphene for the BAs and BSb monolayers [43]). As a consequence, excellent thermoelectric figure-of-merit $ZT = S^2\sigma T / (\kappa_e + \kappa_l)$ can be expected in the BX monolayers. It should be noted that the $\kappa_l$ of BP system is quite high, which could be significantly decreased by using a bilayer structure to enhance the ZT value, as discussed in our previous work.[44] Here we focus on the thermoelectric performance of BAs and BSb monolayers. Figure 3(a) and 3(b) plot the temperature dependent ZT values of the BAs and BSb monolayers, respectively. Compared with those along the armchair direction, we see higher ZT values along the zigzag direction for both p- and n-type carriers, which should be attributed to the lower lattice thermal conductivities discussed above. Due to relatively larger band gaps, the best thermoelectric performance of the Bas monolayer is achieved at a higher temperature of 1300 K compared with that at 700 K for the BSb system. The corresponding ZT values as a function of carrier concentration are plotted in Fig. 3(c) and 3(d), where we find that both systems exhibit better p-type thermoelectric performance along the zigzag direction. At the optimized hole concentration of $5.7 \times 10^{19}$ cm$^{-3}$ ($2.7 \times 10^{19}$ cm$^{-3}$), the highest ZT values of 3.7 (3.3) can be obtained for the BAs (BSb) monolayer, which exceed the target value of 3.0 for the practical applications of thermoelectric materials. Moreover, it is interesting to find that along the armchair direction, the p- and n-type ZT values of BSb are almost identical to each other, which is quite beneficial for fabrication of thermoelectric modules with comparable efficiencies.

In summary, our theoretical work suggests that the graphene-like BAs and BSb monolayers could achieve considerably larger ZT which even exceed the target value of thermoelectric applications. As opposite to that of graphene, such a record high performance originates from the moderate band gap and obviously lower thermal conductivity, which is believed to be rooted in the presence of two different hexagonal sublattices. Experimentally, the BAs and BSb monolayers could be prepared by mechanical cleavage or chemical vapor deposition similar to the isolation of single-layer hexagonal BN.[45−47] In general, the underlying design principle of our theoretical study could be used to find high performance thermoelectric materials in other two-dimensional systems such as layered InP$_3$,[48] PBi,[49] and transition metal dichalcogenides.[50−52]

We thank financial support from the National Natural Science Foundation (Grant Nos. 51772220 and 11574236). The numerical calculations in this work have been done on the platform in the Supercomputing Center of Wuhan University.

**Table I**. Comparison of acoustic phonon group velocities (at 300 K) of BX monolayers.

|     | ZA (km/s) | TA (km/s) | LA (km/s) |
| --- | --- | --- | --- |
| BP  | 3.2 | 7.9 | 13.9 |
| BAs | 1.6 | 6.2 | 10.1 |
| BSb | 1.4 | 4.1 | 7.5  |

**Table II**. The room temperature electronic transport coefficients of $p$-type BX monolayers at the optimized carrier concentration along the zigzag direction.

|     | $n$ ($10^{19}$ cm$^{-3}$) | $S$ ($\mu$V/K) | $\sigma$ ($10^4$ S/cm) | $\kappa_e$ (W/mK) | $S^2\sigma$ (W/mK$^2$) | $\mu$ (m$^2$/VK) |
| --- | --- | --- | --- | --- | --- | --- |
| BP  | 4.9 | 199 | 2.28 | 5.97 | 0.09 | 0.29 |
| BAs | 1.6 | 240 | 3.16 | 8.28 | 0.18 | 1.20 |
| BSb | 1.4 | 251 | 3.29 | 6.53 | 0.21 | 1.66 |

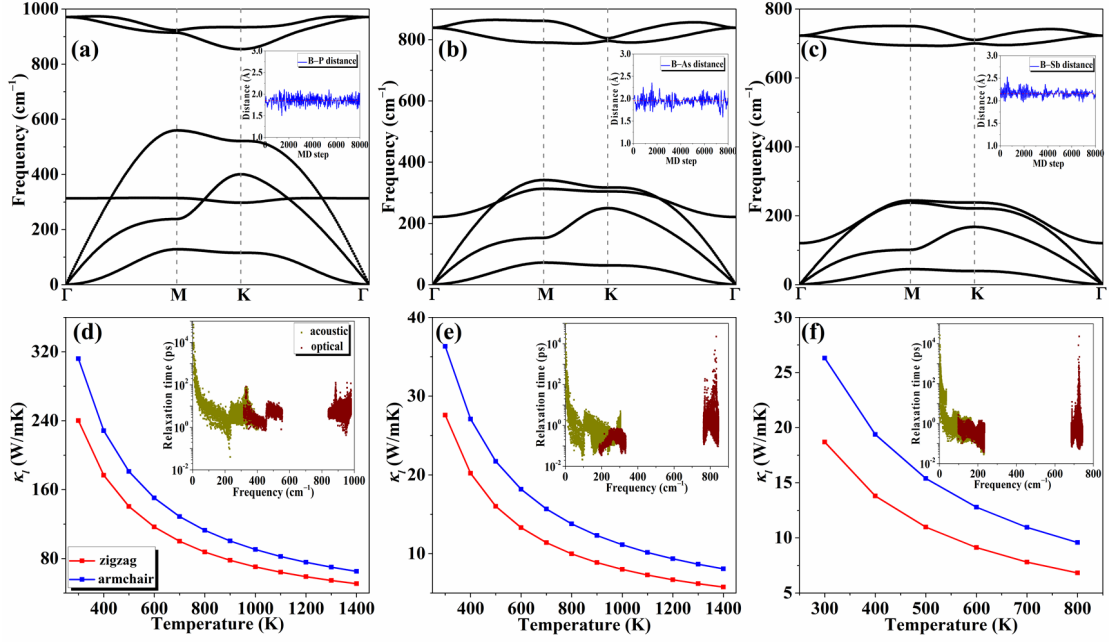

**Figure 1.** The phonon dispersion relations of (a) BP, (b) BAs, and (c) BSb monolayers. The insets of (a), (b), and (c) give the AIMD calculated B−P, B−As, and B−Sb distances at 1400 K, respectively. (d), (e), and (f) respectively shows the temperature dependent lattice thermal conductivities of BP, BAs, and BSb monolayers, and the room temperature phonon relaxation times are indicated in the insets.

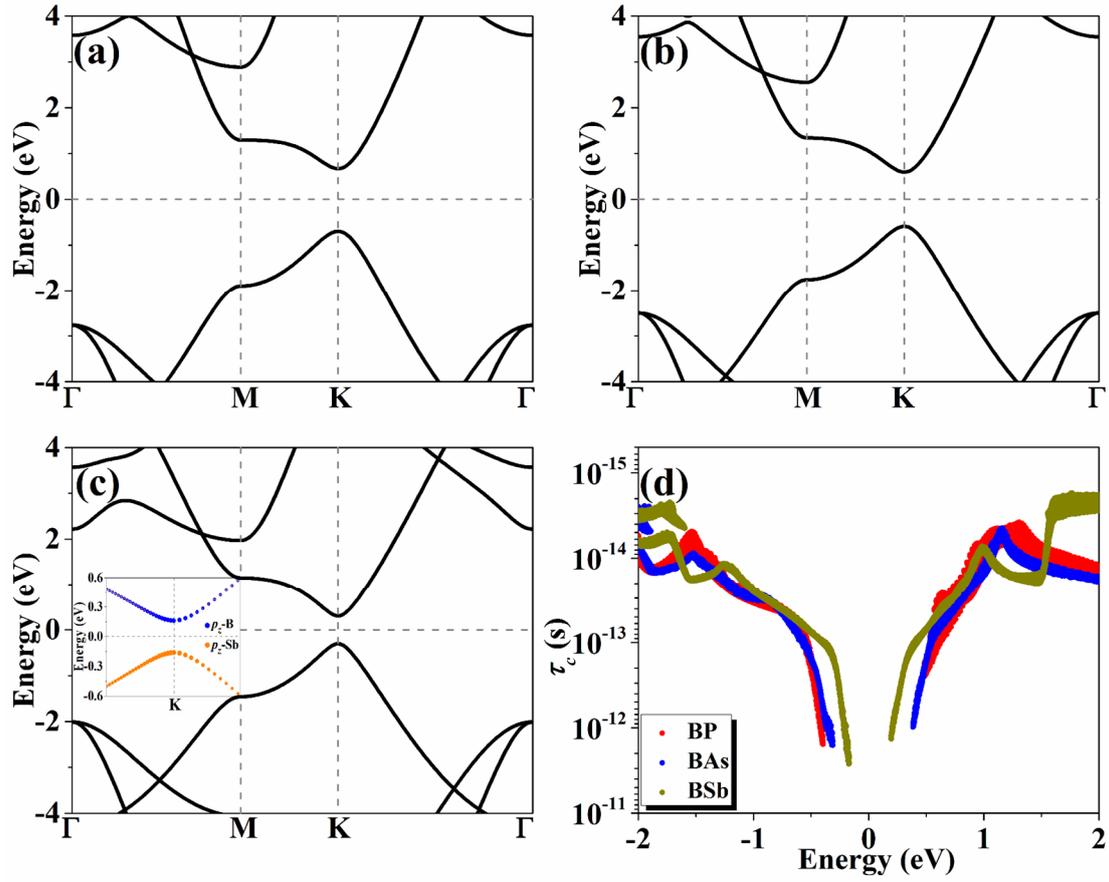

**Figure 2.** The band structures of (a) BP, (b) BAs, and (c) BSb monolayers calculated by using the HSE functional. The inset in (c) shows the orbital-decomposed band structures of BSb system around the Fermi level. (d) is the energy-dependent carrier relaxation time of BX monolayers at 300 K. The Fermi level is at 0 eV.

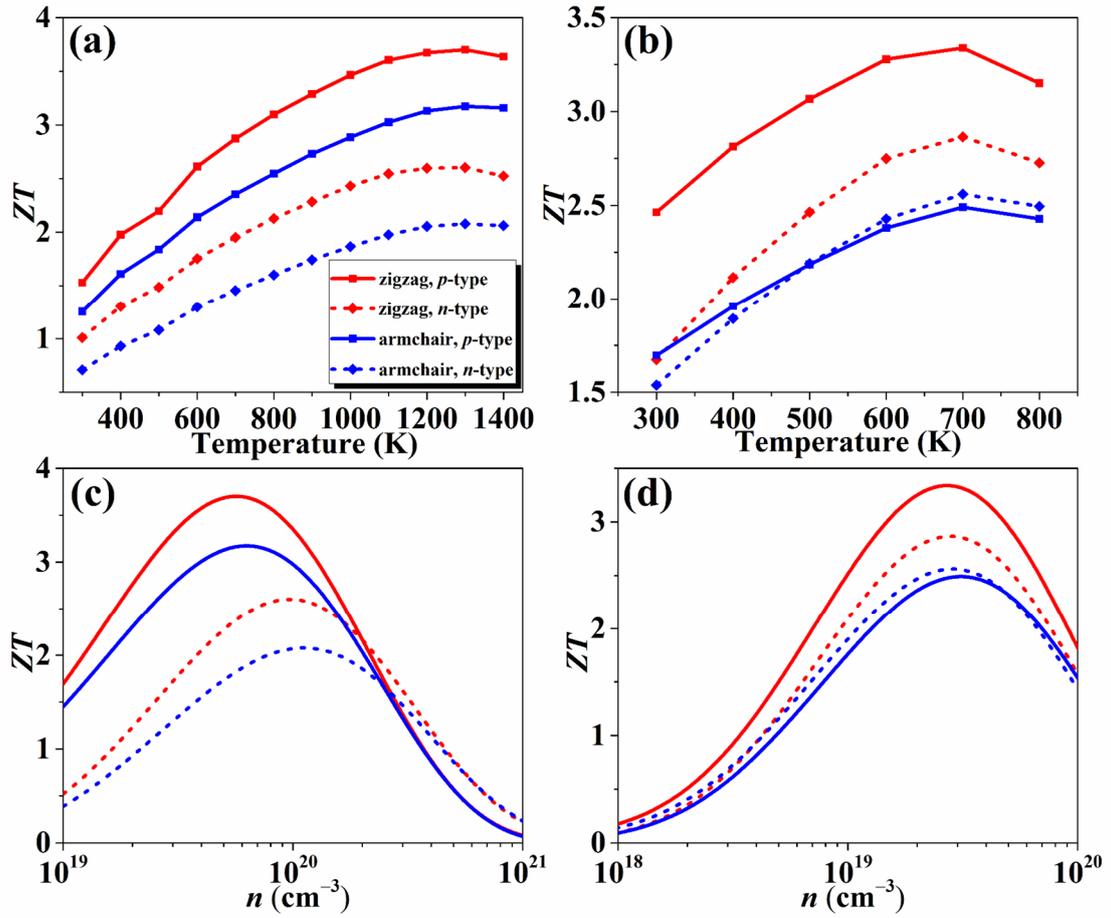

**Figure 3.** The temperature dependent *ZT* values of (a) BAs and (b) BSb monolayers. (c) and (d) are respectively the *ZT* values of BAs at 1300 K and BSb at 700 K, plotted as a function of carrier concentration.